# Usage of infinitesimals in the Menger's Sponge model of porosity

Maria C. Vita[1], Samuele De Bartolo[1], Carmine Fallico[1], Massimo Veltri[1]

[1]University of Calabria, Department of Soil Conservation,
Cubo 42B V piano, P. Bucci, Rende (CS) 87036 - Italy
vita@dds.unical.it, samuele.debartolo@unical.it, fallico@dds.unical.it, veltrim@dds.unical.it

**Abstract**

The present work concerns the calculation of the infinitesimal porosity by using the Menger's Sponge model. This computation is based on the grossone theory considering the pore volume estimation for the Menger's Sponge and afterwards the classical definition of the porosity, given by the ratio between the volume of voids and the total volume (voids plus solid phase). The aim is to investigate the different solutions given by the standard characterization of the porosity and the grossone theory without the direct estimation of the fractal dimension. Once the utility of this procedure had been clarified, the focus moves to possible practical applications in which infinitesimal parts can play a fundamental role. The discussion on this matter still remains open.

## 1. Introduction

When dealing with natural porous materials, the knowledge one can achieve is limited. This is often due to the fact that the inner nature of certain systems cannot be investigated. For this reason, especially when it comes to infinite and infinitesimal processes, one usually tends to approximate because it is not possible to fully describe them. In some case, however, it is important to get a precise estimation of volumes, whatever small or large they could be.

In general the characterization of porous media is based on particular mathematical models, in which the role of the voids is strictly related to the solid matrix properties [1, 2]. For example, natural porous media are physically not easy to study. They are usually schematized by means of a specific deterministic model, like the Menger's Sponge [3].

It is well known that at infinity the volume of the Menger's Sponge is equal to zero [3]. But we don't know how far infinity is nor we know the volume at $\infty$-1, $\infty$-2, $\infty$-3,.... And the difference between two steps can be crucial under certain circumstances. An easy way to understand the importance of the matter is to think about the definition, for example, of the low water content in the unsaturated soils [4] where infinitesimal parts can be relevant for large agricultural systems [2].

In particular, natural soils can be described through the fractal geometry. In the framework of the transport in porous media as well as in the water retention curves, that characterize the flow processes in the unsaturated zone, many works have already proved the likelihood of this passage by using the Menger's Sponge model [5, 6, 7, 8].

The Menger's Sponge is a mathematical set, whose geometrical properties are defined over the whole observation scales in an iterative way. Anyway the precision concerning the infinitesimal volumes and areas for natural phenomena is influenced by the cut-off scaling procedure, by which our measurements are performed. This last aspect can be relevant if we have the necessity to explore some part of it with more precision [9].

## 2. Water Retention Curves Fractal Models

The water retention curves (WRC) represent the variation of the water content, $\theta$, in the soil with the matrix potential, $\psi$, or, more commonly, the hydraulic head, h.

The amount of water content in the soil, $\theta(h)$, showing a decreasing trend with increasing of hydraulic head, influences many processes as the diffusion of nutrients to plant roots, the soil temperature, the velocity with which the water and the solutes move through the root zone, etc [10].

Generally these curves play a fundamental role in the description and modeling of the flow and the mass transport in the unsaturated soil. Taking into account that the water behavior depends on pore network geometry, in the last years many WRC fractal models have been studied. Among the most well-known of these it is possible certainly to include the Tyler and Wheatcraft (TW) model [11], the Rieu and Sposito (RS) model [12, 13] and the Pore Solid Fractal (PSF) model [14]. This last represents the most general fractal model, while the TW and RS models can be considered particular cases of it; therefore starting from the PSF model it is possible to obtain the other two. Following, the PSF model will be introduced and with simple limits considerations the TW and RS will be then derived.

With reference to the PSF model, assuming that both the solid and pore phases exhibit self-similarity properties, the soil can be characterized in terms of fractal geometry and described by fractal scaling laws [15]. Bird *et al*. [16] developed a general fractal model that can describe either soil structure or the symmetry between the solid and pore phases. Specifically, using general iterative and Menger's Sponge-like descriptive procedures, the authors related two distribution phases to the bulk volume of the pore and solid phases via a power law. First introduced by Perrier *et al*. [14], who called it the pore-solid fractal (PSF), this model derives soil water content from the following expression [16, 17]:

$$\theta(h) = \Phi - \frac{p}{p+s}\left[1 - \left(\frac{h}{h_{min}}\right)^{D_f - E}\right] \qquad h_{min} \leq h \leq h_{max} \qquad (1)$$

where $\Phi$ is the total porosity, $h_{min}$ and $h_{max}$ are the pressure head values at which the largest and the smallest pores explicitly defined within the PSF become unsaturated, and p and s are the pore and solid phase fractions at each scale. Given a three-dimensional Euclidean domain (E=3) and defining A [18] as:

$$A = \frac{p}{p+s}, \qquad (2)$$

yields the following equation:

$$\theta(h) = \Phi - A\left[1 - \left(\frac{h}{h_{min}}\right)^{D_f - 3}\right]. \qquad (3)$$

Assuming that soil pores can be fully saturated [19], equation (1) can be re-written [16] as:

$$\theta(h) = (\theta_S - A) + A\left(\frac{h}{h_{min}}\right)^{D_f - 3}, \qquad (4)$$

where $\theta_S$ is the saturated water content ($\theta_S = \Phi$).

In the Bird *et al.* model [16], the value of the parameter A is variable between the saturated water content ($\theta_S$) and 1, that is $\theta_S \leq A \leq 1$. It is opportune to observe that the TW and RS models previously introduced are limited cases of the more general Bird *et al.* model [16]. In fact, assuming A=$\theta_S$, i.e., when the solid phase is predominant (p→0), equation (1) describes a structure that tends toward a fractal porous mass. In this case, the general PSF model and therefore the water retention function can be expressed in terms of the TW model. Similarly, assuming A=1 in equation (4), i.e., when the solid phase is zero, the general Bird *et al.* model [16] and therefore the water retention function assumes the form of the RS model. Consequently, the different water retention models can be regarded as being specific cases of the Bird *et al.* model [16] represented in equation (1). The TW and RS models are discussed hereinafter.

In Figure 1 the WRC is represented according to PSF model. Specifically the trend $\theta(h)$ is shown (h in cm) and the scaling behavior of the equation (4), i.e. $\log((\theta + A - \theta_s)/A)$ versus $\log(h_{min}/h)$. This last shows a linear behavior, with the parameters conditions $\theta_s = 0.5$, A=0.45 and $h_{min}=1$. This trend is however theoretic, while in real cases it results different; i.e. it is characterized by two scaling regions, namely a bi-modal curve, in correspondence of high and low water contents respectively. This aspect is shown in Figure 2 referred to a real case (sandy-loam soil). In the same figure these two scaling regions are shown; for each of them two fractal dimensions can be estimated. In any case the assessment of the water content values for the second range (low values) results very difficult to determine on a laboratory scale. Concerning porosity, equation (1) can be integrated through the equation (31) for opportune *n* values.

As anticipated, the PSF model is a general one and represents the basis to develop other models. In order to complete the relationships previously introduced, i.e. Tyler and and Wheatcraft (TW) [11] and Rieu and Sposito (RS) [12, 13], hereinafter we define more precisely the framework concerning the soil pore structure as a fractal representing the pore and solid phases for both models [4]. The principle they follow is to relate the hydraulic properties of soils to the geometrical ones of the pore space and to consider a simple capillary model to combine the WRCs with a pore-size distribution [4]. This way the value of the volumetric water content, θ, at the pressure head, h, is set as that of the porosity $\Phi \leq l$, due to the pore of size smaller than the characteristic dimensionless $l=\alpha/h$, where the characteristic length of capillarity $\alpha$ is a dimensional constant measured in centimeters. Tyler and and Wheatcraft and Rieu and Sposito models are then obtained a mass fractal pore-size distribution.

The TW model [11] is based on the Menger's Sponge and does not consider a lower cut-off of scale for a minimum pore-size $l_{min}$ tending to zero. Starting from the following relationship:

$$[\Phi \leq l] = \Phi \left(\frac{l}{l_{max}}\right)^{E-D_f}, \tag{5}$$

where $l_{max}$ is the maximum pore-size, E is the Euclidean dimension and $D_f$ is the fractal dimension, by applying the Laplace law we obtain:

$$\theta(h) = \left[\Phi \leq \frac{\alpha}{h}\right] = \Phi \left(\frac{h}{h_{min}}\right)^{D_f-E}, \tag{6}$$

where $h_{min}$ is the water potential usually known as air-entry value [20], inversely proportional to $l_{max}$. The final analytic expression given by the TW model for θ(h) is:

$$\theta(h) = \theta_s \left(\frac{h}{h_{min}}\right)^{D_f-E}, \tag{7}$$

where $\theta_s$ is the saturated water content.

Rieu and Sposito modeled the WRCs on the hypothesis that a fractal mass described by a two-phase (pore ad solid) geometrical model is characterized by a finite minimum cut-off limit of scale, so that $l_{min} \neq 0$. The porosity, Φ, is therefore expressed as follows:

$$[\Phi \leq l] = \Phi_{max} - 1 + \left(\frac{l}{l_{max}}\right)^{E-D_f}, \tag{8}$$

where $\Phi_{max}$ is the porosity at saturated conditions. Here too it is possible to apply the Laplace law, obtaining:

$$\theta(h) = \left[\Phi \leq \frac{\alpha}{h}\right] = \Phi_{max} - 1 + \left(\frac{h}{h_{min}}\right)^{D_f-E}, \tag{9}$$

and, ultimately:

$$\theta(h)=\theta_s\text{-}1+\left(\frac{h}{h_{\min}}\right)^{D_f\text{-}E}. \tag{10}$$

As equations (1), (3) and (6) clearly show, the role of the porosity can be relevant for a correct characterization of the water retention curves. For this reason the evaluation of the volume of the infinitesimal pores can be of great interest for a better assessment of the porosity, whatever scaling process is considered and, consequently, without taking into account the fractal dimension.

In the following paragraphs we introduce different methods for the evaluation of infinitesimal volumes according to the classical Menger's Sponge model and to the grossone theory.

### 3. Evaluating infinitesimal volumes and areas: the case of Menger's Sponge and Sierpiński Carpet

The Menger's Sponge is a 3D fractal object obtained as follows: starting from a solid cube of volume normalized to 1, by dividing each dimension by three 27 sub-cubes are obtained and on each face of the cube there are nine sub-cubes. The central one and the six central sub-cubes on each face must be removed, having left only 20 sub-cubes. This procedure can then be reiterated (see Figure 3a) [21]. Considering $n$ iterations, the geometrical characteristics of the Menger's Sponge are represented by the following expressions [9]:

$$N_n = 20^n, \tag{11}$$

$$L_n = \left(\frac{1}{3}\right)^3 = 3^{-n}, \tag{12}$$

$$V_n = L_n^3 N_n = \left(\frac{20}{27}\right)^n, \tag{13}$$

$$A_n = L_n^2 N_n = \left(\frac{8}{9}\right)^{n-1}, \tag{14}$$

where $N_n$ is the number of filled boxes, $L_n$ is the length of the side of a hole (or a filled part), $V_n$ is the fractional volume after the $n$th iteration and $A_n$ represents the area of the solid cubes of the front sections of the Menger's Sponge, which corresponds to the two-dimensional fractal scheme of the Sierpiński's Carpet (Figure 3b). From the theoretical point of view, the classical Menger's Sponge model has its limits because of the assumption of the scheme itself, which implies that at infinity the volume of the Sponge is zero and, in the case of the Sierpiński's Carpet, its area is infinite. Nevertheless in many practical applications, for systems modeled by means of the Menger's Sponge, the fractal dimension plays a fundamental role to characterize only the scaling processes related to the mathematical model used to describe a natural phenomenon [22]. The so-calculated fractal dimension represents an assessment and doesn't coincide with the fractal dimension in the Haussdorff sense but only with the box-counting dimension [23, 24, 25]. For this reason it must be determined over the cut-off scale lengths [22]. For deterministic cases, like the Menger's Sponge, the above dimensions are coincident [23, 25] and don't correspond to the topological dimension [23, 24]. Therefore the fractal dimension of the Menger's Sponge is defined through the following limit process:

$$D_f = -\lim_{n\to\infty} \frac{\ln N_n}{\ln L_n} = \frac{\ln 20}{\ln 3} = 2.726833028\ldots, \tag{15}$$

where *n* indicates the number of iterations performed and $N_n$ and $L_n$ have the meanings mentioned above.

Expression (15) is representative of a limit process. But limits cannot give a precise estimation; they just show the behavior of a function when the independent variable tends to a specific value within the cut-off scale lengths [22]. What information can we get beyond the limits of the scaling process if they don't show precise cut-off points? The value of *infinity* cannot be defined in processes like the ones above described: it can be one million, ten millions, a hundred millions; but one can always reach one step further. How can we determine that, considering a particular phenomenon? One can be interested in estimating a quantity very precisely if there is a more extensive, unlimited scaling trend. It can even be crucial to do so. Therefore, new and more appropriate tools are necessary to improve the precision of the estimation of such infinitesimal quantities. And these tools must be powerful enough to estimate also infinite quantities. On this matter it results of great importance the work of Sergeyev [26, 27, 28, 29, 30, 31, 32] with the *grossone* theory.

This theory is based on the principle that "the part is less than the whole" and opens to the new scenario of considering infinite and infinitesimal numbers the same way as finite quantities, so that they can be treated numerically [9]. This way it becomes possible to appraise the difference between ∞ and ∞-1.
Even though infinite and infinitesimal numbers can now be treated the same way as the finite ones, they still remain different. So a new unit of measure is necessary to quantify them. This unit of measure is precisely the grossone (①), which is infinite and equal to the number of elements of the set ℕ of natural numbers [9]. It is important to underline that ℕ is still composed of the same elements. The distinction stands in a different system used to express ℕ: ① is an infinite unit of measure. For further informations about the grossone one can refer to the work of Sergeyev [26, 27, 28, 29, 30, 31, 32].

The aim of this paper is only to focus on the estimation of volumes by means of the grossone for porous media modeled by means of the Menger's Sponge and in particular for the estimation of the porosity in terms of grossone theory. Nevertheless it is necessary to briefly introduce the subject.

The grossone theory is based on the following three postulates [9]:

- **Postulate 1**: There exist infinite and infinitesimal objects but human beings and machines are able to execute only a finite number of operations.
- **Postulate 2**: People do not tell what are the mathematical objects they deal with; they just will construct more powerful tools that will allow to improve their capacities to observe and to describe properties of mathematical objects.
- **Postulate 3**: The principle "The part is less than the whole" is applied to all numbers (finite, infinite and infinitesimal) and to all sets and processes (finite and infinite).

The third postulate is particularly important for this study because allows to treat infinite numbers the same way of finite numbers [9].

Another important standing point, as said before, is that the set $\mathbb{N}$ of natural numbers has $①$ elements. Given the definition of infinite sequence as "a function having as the domain the set of natural numbers $\mathbb{N}$, and as the co-domain a set A", we have that any sequence having $\mathbb{N}$ as the domain has $①$ elements. This means that any subsequence is made of $①$ elements or less. That is to say that any sequential process can have maximum $①$ steps [9]. Besides, $①$ has both ordinal and cardinal properties [9].

These are the most important things to set out for our purpose. It is now possible to calculate the volumes of the Menger's Sponge for $①$ iterations, which are as follows [9]:

$$N_{1,n} = 20^{n-1}, \tag{16}$$

$$L_{1,n} = 3^{-(n-1)}, \tag{17}$$

$$V_{1,n} = L_{1,n}^3 N_{1,n} = \left(\frac{20}{27}\right)^{n-1}, \tag{18}$$

For $n=①$:

$$V_{1,①} = \left(\frac{20}{27}\right)^{①-1}. \tag{19}$$

For $n=①-1$:

$$V_{1,①-1} = \left(\frac{20}{27}\right)^{①-2} \tag{20}$$

and so on. Ultimately, for $1 \leq k \leq n \leq ①+k-1$ we have:

$$V_{k,n}=L_{k,n}^3 N_{k,n}=\left(\frac{20}{27}\right)^{n+k-2}, \tag{21}$$

where *n* is the number of iterations, *k* is the initial level of the Menger's Sponge from which we start the process and $N_n$ and $L_n$ have the meaning mentioned above. This represents the volume of the solid phase of the Menger's Sponge according to the grossone theory.

In terms of areas, following the Sierpiński's Carpet scheme, the following equation furnishes the corresponding equation of (14) in terms of grossone for $1 \leq k \leq n \leq ①+k-1$:

$$A_{k,n}=L_{k,n}^2 N_{k,n}=\left(\frac{8}{9}\right)^{n+k-2}. \tag{22}$$

## 4. Evaluating infinitesimal porosity

It is straightforward to think that, when dealing with porous media, the first and most important property to estimate is the *porosity* [33, 34].

Physically, the porosity represents the voids within a certain solid matrix. It is set as the ratio between the volume of voids, $V_V$, and the total volume, $V_T$, given by the sum of pore and solid phase ($V_S$). So in general its relationship is expressed by the following ratio:

$$\Phi=\frac{V_V}{V_T}=\frac{V_V}{V_V+V_S}. \tag{23}$$

Considering the classical process based on the Menger's Sponge, the porosity, $\Phi$, is given by the following scheme:

$$V_S(n)=\left(\frac{20}{27}\right)^n \tag{24}$$

is the volume of the solid phase. With regard to the volume of the pore phase, the relationship is expressed by the complement to 1 of the volume of the solid phase, starting from a full solid volume:

$$V_V(n)=1-\left(\frac{20}{27}\right)^n, \tag{25}$$

where *n* is the number of iterations. By substituting (24) and (25) in (23), we obtain:

$$\Phi(n)=\frac{V_V(n)}{V_V(n)+V_S(n)}=\frac{1-\left(\frac{20}{27}\right)^n}{1-\left(\frac{20}{27}\right)^n+\left(\frac{20}{27}\right)^n}=1-\left(\frac{20}{27}\right)^n \tag{26}$$

The right term of equation (26) represents the porosity of a porous medium in terms of the Menger's Sponge, namely:

$$\Phi(n)=1-\left(\frac{20}{27}\right)^n. \tag{27}$$

### 4.1 Porosity of the Menger's Sponge: fractal approach

We have until now considered the most classical of the Menger's Sponge constructions. However it is not the only one. Turcotte [35], for instance, used to make up the Sponge also starting from solid cubes characterized by specific density, $\rho_0$, and size, $r_0$. This way the first-order Sponge is formed by a number of cubes of size $r_0$ in order to achieve a size $r_1=3r_0$ and contains 20 solid zero-order cubes. Therefore the porosity is obtained as follows:

$$\Phi_1 = \frac{7}{27}, \tag{28}$$

and the density is:

$$\rho_1 = \frac{20}{27}\rho_0. \tag{29}$$

Proceeding with the iterations, the second-order Sponge is characterized by a size $r_2=9r_0$ and is made up of 400 solid cubes of size $r_0$ and density $\rho_0$. So we have that the porosity is:

$$\Phi_2 = \frac{329}{729}, \tag{30}$$

and the density is:

$$\rho_2 = \frac{400}{729}\rho_0. \tag{31}$$

For the $n$th-order Sponge we have the following expressions:

$$\Phi = 1 - \left(\frac{r_0}{r_n}\right)^{3-\ln 20/\ln 3}, \tag{32}$$

$$\frac{\rho_n}{\rho_0} = \left(\frac{r_0}{r_n}\right)^{3-\ln 20/\ln 3}. \tag{33}$$

In general, in terms of fractal dimension the porosity and the density of an $n$th-order Menger's Sponge are given by the following equations:

$$\Phi = 1 - \left(\frac{r_0}{r}\right)^{3-D_f}, \tag{34}$$

$$\frac{\rho_n}{\rho_0} = \left(\frac{r_0}{r}\right)^{3-D_f}, \tag{35}$$

where $r$ is the linear dimension of the volume considered.

It is now important to remark a few things. Equation (27), even though follows the same principle of the expression given by Katz and Thompson [36], as said by Turcotte [35], is not a power-law (fractal) relation. For a direct comparison between equation (27) and a power-law (fractal) relation it is necessary to consider the density of the $n$th-order Menger's Sponge, by which it is possible to define the exponent as a function of the fractal dimension, which is obtained for structural matrix like soil aggregates [36, 35]. Anyway, this last aspect, as shown below, doesn't concern the porosity in terms of grossone, they being different relationships because Turcotte's depends on the fractal dimension, $D_f$.

### 4.2 Porosity of the Menger's Sponge: the grossone approach

According to the grossone theory [9], it is now possible to estimate the geometrical properties of an object even for an infinite number of iterations. Obviously, this also includes the porosity, which, in the case of the Menger's Sponge, can be calculated as follows. For the volume of the solid phase the expression is:

$$V_S(n) = \left(\frac{20}{27}\right)^{n+k-2} \tag{36}$$

where $n$ is the number of iterations, $k$ is the level at which the analysis is started.

With regard to the volume of the pore phase, the relationship is expressed by the complement to 1 of the volume of the solid phase:

$$V_V(n) = 1 - \left(\frac{20}{27}\right)^{n+k-2}. \tag{37}$$

By substituting (28) and (29) in (23), we obtain:

$$\Phi(n) = \frac{V_V(n)}{V_V(n)+V_S(n)} = \frac{1-\left(\frac{20}{27}\right)^{n+k-2}}{1-\left(\frac{20}{27}\right)^{n+k-2}+\left(\frac{20}{27}\right)^{n+k-2}} = 1 - \left(\frac{20}{27}\right)^{n+k-2} \tag{38}$$

The right term of the equation (30) represents the porosity of a porous medium in terms of the Menger's Sponge and the grossone theory, namely:

$$\Phi(n) = 1 - \left(\frac{20}{27}\right)^{n+k-2} \tag{39}$$

which is defined for $1 \leq k \leq n \leq ① + k - 1$.

In general equation (39) allows, unlike equation (27), to perform a local evaluation starting from the $k$th iteration as if it was an enlargement procedure.

### 5. Conclusions and open discussions

The grossone theory offers the possibility to investigate the infinity and the infinitesimal for a variety of systems [9]. The framework of the grossone theory can be extended to the modeling of porous media, particularly for the definition of the porosity. This parameter is very important for many practical applications because it affects, for instance, flow and solute transport in groundwater or other porous media, where the support scale is relevant [1, 4]. Generally, as regards pore formation and soil aggregates, the porosity is strictly related to the fractal dimension through power-law fractal relationships, where the fractal dimension itself is defined only over limited cut-off scale lengths [22]. This aspect, however interesting, constitutes a restriction because some physical processes, like the water retention curves, show an unlimited scaling range, particularly for infinitesimal volume parts [4]. The interesting thing of it is due to the definition of the volumes of low water contents (or other physical agents), for which the same scaling process shows an infinite scaling behaviour that is not directly calculable but by onerous and time-consuming procedures [37]. In this context, the grossone theory enables to achieve a higher precision to determine the infinitesimal volumes if models like the Menger's Sponge are used as a basis to describe the porous medium. Unlike the classical definitions for the porosity of the Menger's Sponge, based on power-law fractal relationships in which it is possible to define the exponent as a function of the fractal dimension, by means of the grossone theory it is possible to overcome the restriction due to the assessment of the same dimension. At least for what concerns the scaling processes with an unlimited trend, as reported in Figure 2. This last aspect is strictly related to the fractal characterization of the water retention curves, where the low water content plays a predominant role at different measurement scales, whether local or spatial [4, 38]. Anyway the correspondence between equation (1) and equation (27) is dual, namely in terms of total porosity offers the possibility to investigate the low water content region by using appropriate $n$ values, without considering the lower scale length of this region exploring more significant parts. Therefore by means of the grossone theory it is possible to achieve a more precise estimation of the porosity, $\Phi$, as expressed in equation (39), which allows to perform a local evaluation starting from the $k$th iteration as if it was an enlargement procedure. While, adopting the same fractal PSF model, we can also define the low cut-off limit of the water retention curves as precisely as we do with $\Phi$. This means a better quantification of the exploitable water content in the unsaturated zone, locally and spatially.

**Figure captions**

Figure 1: PSF model for a theoretical WRC: this plot is obtained with the limit conditions $\theta_s$=0.5, A=0.45 and $h_{min}$=1. In the frame box is reported the straight line scaling behavior in the logarithm scale.

Figure 2: PSF model of a real WRC: this plot is obtained for the limit conditions $\theta_s$=0.501, A=0.45 and $h_{min}$=0.01. In the frame box is reported the typical bi-modal scaling behavior of real cases.

Figure 3: a) Menger's Sponge iteration scheme; b) Sierpiński Carpet scheme.

**Figures**

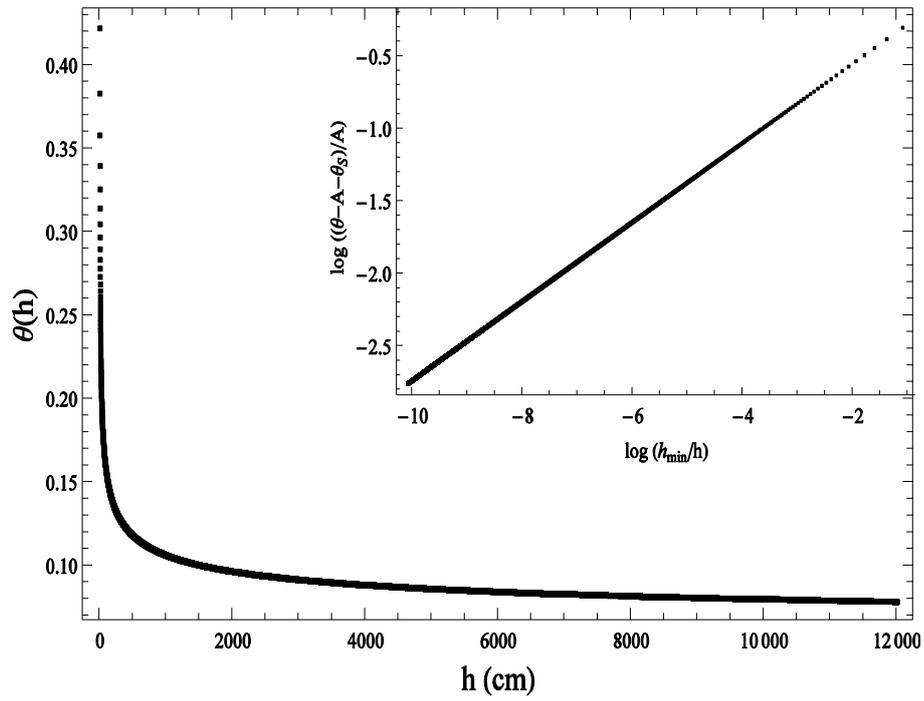

Figure 1

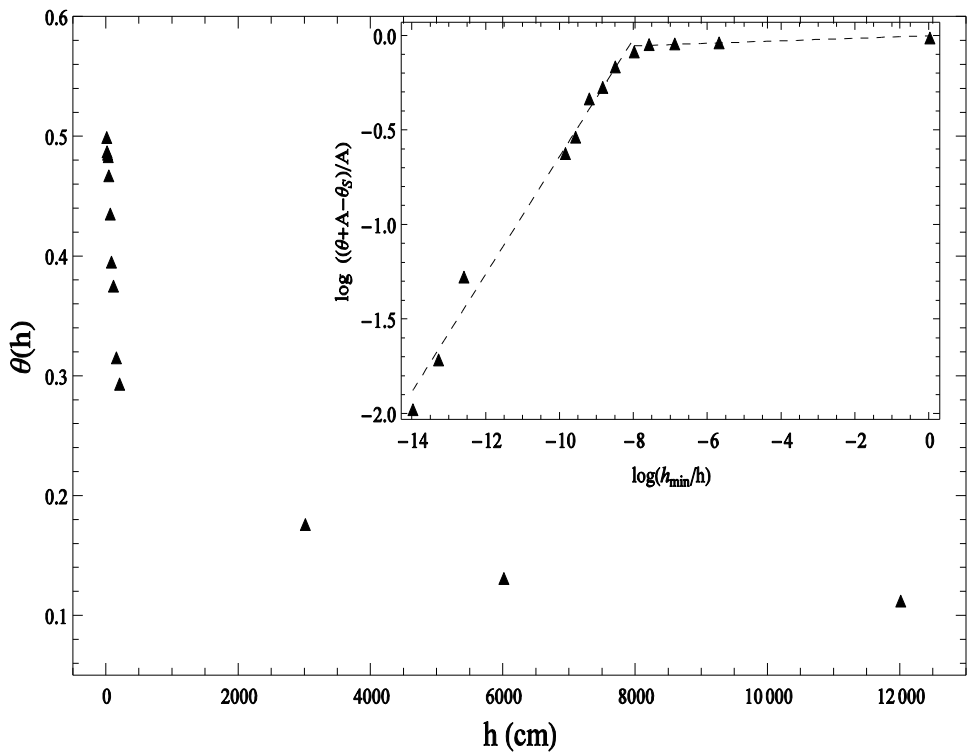

Figure 2

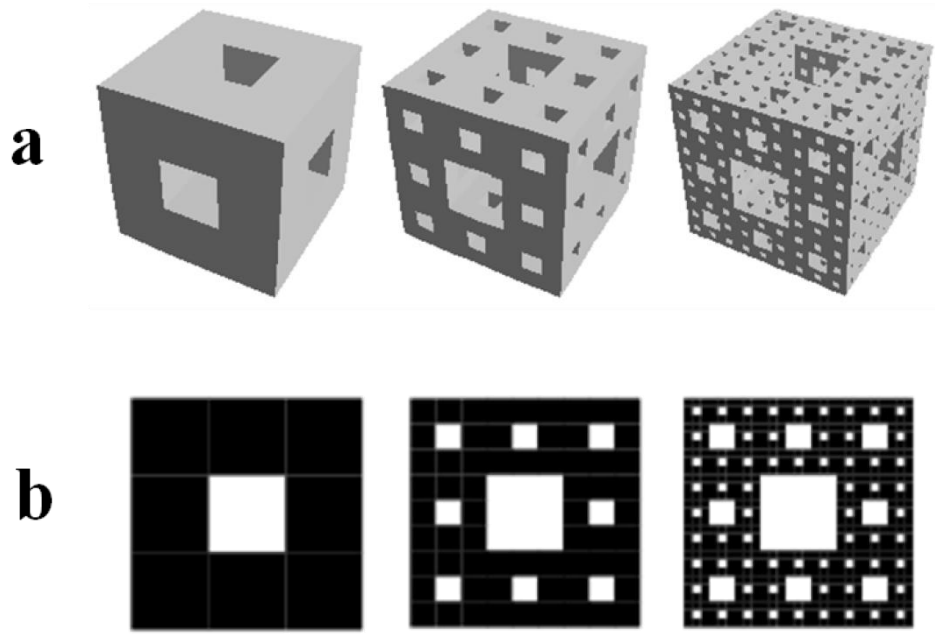

Figure 3